\documentclass[prl,showpacs,superscriptaddress,twocolumn]{revtex4}

\usepackage{graphicx}
\usepackage{color}
\usepackage{latexsym}
\usepackage{amsmath}
\usepackage{amsthm}
\usepackage{amsfonts}
\usepackage{amssymb}
\usepackage{bm}

\newcommand{\rem}[1]{}
\newcommand{\refe}[1]{(\ref{#1})}

\newcommand{\refE}[1]{Eq.~(\ref{#1})}
\newcommand{\beq}{\begin{equation}}
\newcommand{\eeq}{\end{equation}}
\newcommand{\beqa}{\begin{eqnarray}}
\newcommand{\eeqa}{\end{eqnarray}}
\newcommand{\eps}{\varepsilon}

\begin{document}
\title{
Multiple Andreev reflections in hybrid multiterminal junctions
}

\author{M. Houzet}

\affiliation{SPSMS, UMR-E 9001, CEA-INAC/UJF-Grenoble 1, F-38054 Grenoble, France
}

\author{P. Samuelsson}
\affiliation{Division of Mathematical Physics, Lund University, Box 118, S-221 00 Lund, Sweden}

\date{\today}

\begin{abstract}
We investigate theoretically charge transport in hybrid multiterminal
junctions with superconducting leads kept at different voltages. It is
found that multiple Andreev reflections involving several
superconducting leads give rise to rich subharmonic gap structures in
the current-voltage characteristics. The structures are evidenced
numerically in junctions in the incoherent regime.

\end{abstract}

   		
\maketitle

The mechanism of charge transfer across an interface between a normal
(N) conductor and a superconductor (S) at subgap energies is Andreev
reflection (AR) \cite{andreev:1964}. In an AR an electron incident
from N is retroreflected as a hole, creating at the same time a Cooper
pair in S. In SNS-junctions subjected to a voltage bias $V$,
particles can undergo several ARs at the NS-interfaces, gaining energy
$eV$ at each traversal of the junction, and escape out in the leads at
energies above the superconducting gap $\Delta$. This process of
multiple Andreev reflections \cite{tinkham:1982,octavio:1983} (MAR) is
responsible for the charge transport at voltages $eV<2\Delta$ and
gives rise to a subharmonic gap structure (SGS) at $eV=2\Delta/n$ ($n$
integer) in the current-voltage characteristics.

Almost thirty years after its theoretical description, MAR still
attracts a lot of interest. In recent years, the strongly
non-equilibrium electron distribution caused by MAR was measured in
metallic diffusive junctions~\cite{pierre:2001}. An enhanced shot
noise, due to the multiparticle character of MAR-transport, was
experimentally demonstrated in tunnel \cite{Diel:1997}, metallic
\cite{metalnoise}, and atomic point contact \cite{Cron:2001}
junctions. Superconducting junctions based on new materials and
nanoscale systems such as carbon nanotubes \cite{buitelaar:2003},
semiconductor nanowires~\cite{xiang:2006}, and graphene flakes
\cite{heersche:2007} has allowed for an investigation of the interplay
of MAR and resonant transport, charging effects and Kondo physics, as
well as MAR-transport of Dirac electrons.

While most experimental and theoretical investigations of MAR have
concerned two-terminal structures, multiterminal geometries
with {\it all} superconducting leads have also been studied.
Phase dependent MAR-transport in SNS-interferometers was investigated 
experimentally in a diffusive conductor~\cite{kutchinsky:1999} 
and theoretically in a single mode junction~\cite{lantz:2002}.
In an incoherent three-terminal junction, the current cross-correlation 
between S leads was predicted to be strongly enhanced due to
MARs~\cite{duhot:2009}.

\begin{figure}
\includegraphics[width=0.9\linewidth]{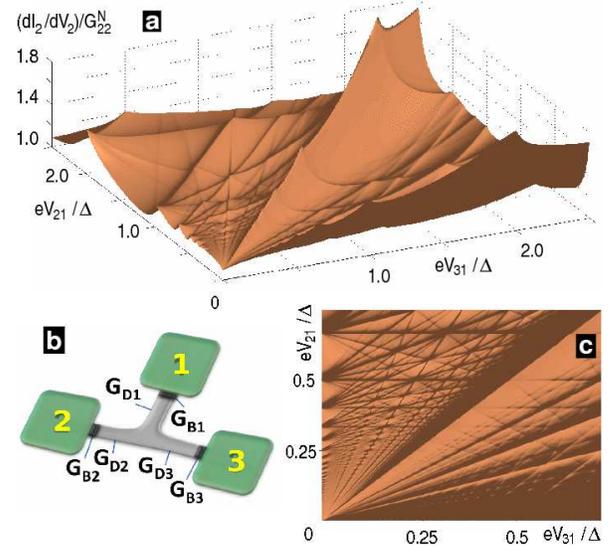}
\caption{ (a) The differential conductance $G_{22}=\partial
I_2/\partial V_2$ normalized with its normal state value as a function
of bias voltages $V_{21}$ and $V_{31}$ in a symmetric
three-superconducting-terminal junction in the incoherent regime. The
conductances $G_{B\alpha}=(2e^2/h)N\Gamma$ with $\Gamma=0.7$ and negligible
resistances $1/G_{D\alpha}$ of the diffusive regions.  The temperature
$kT=0.01\Delta$. (b) Geometry of the junction with superconducting
terminals $\alpha=1,2,3$, barrier conductances $G_{B\alpha}$ and
conductances $G_{D\alpha}$, of the diffusive regions shown.  (c)
Top view of low voltage region of (a).}
\label{fig1} 
\end{figure}

In all these works however, only one bias voltage was applied between
the different leads \cite{Pierre}. In this article, we show that a
much richer picture for SGS and MAR-transport manifests in
multiterminal structures with arbitrary bias voltages. Specifically,
we address a three-superconducting-terminal structure and predict SGS
when (see Fig. \ref{fig1})
\begin{subequations}
\beqa
pV_{21}+qV_{31}&=&2\Delta/e, \,\hspace{0.5cm} \mathrm{or}
\label{eq:MAR-SSS1}
\\
pV_{21}+qV_{31}&=&0.
\label{eq:MAR-SSS2}
\eeqa
\label{eq:MAR-SSS}
\end{subequations}
Here, $p$ and $q$ are integers and $V_{21}=V_2-V_1$
(resp.~$V_{31}=V_3-V_1$) is the bias voltage between leads 2 and 1
(resp.~3 and 1).  This prediction is quite general and could thus be
tested in a broad range of hybrid systems.

In the following, we first provide a physically intuitive description
of the new SGS features in multiterminal junctions with all
superconducting leads.  Thereafter, we present a detailed
investigation of the SGS and MAR-transport for junctions in the
incoherent regime.  This is then contrasted with the result for
multiterminal structures having both normal and superconducting leads.

\begin{figure}
\includegraphics[width=0.9\linewidth]{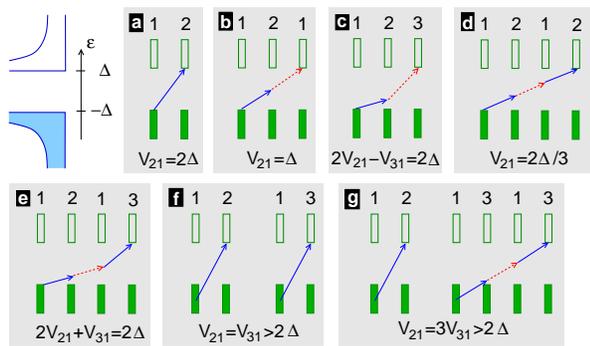}
\caption{Examples of elementary processes determining the SGS lines
(see text). The filled (empty) rectangles represent filled (empty)
states in the gapped energy spectrum of the leads (upper left
fig.). Full (dashed) arrows represent an electron (hole) propagating
between two leads.  The slope of each arrow is set by the bias
voltages.  An AR takes place at the intersection of full and dashed
arrows.  }
\label{fig2} 
\end{figure}

Let us first discuss the SGS given by \refE{eq:MAR-SSS1}. As in
standard two-terminal junctions, the SGS occur at bias voltages for
which new transport processes in energy space become possible, taking
particles from below to above the superconducting gap. In the absence
of AR, direct transmission of an electron through the structure takes
place at $|V_{\alpha\beta}|>2\Delta/e$
($V_{\alpha\beta}=V_\alpha-V_\beta$); the process is illustrated in
Fig.~\ref{fig2}a.  This determines the SGS lines: $V_{21}=\pm
2\Delta/e$, $V_{31}=\pm 2\Delta/e$, and $V_{21}-V_{31}=\pm
2\Delta/e$. Formally, these can be interpreted as the usual MAR features
with $n=1$.

Consider then a process where an electron injected from lead $\alpha$
is Andreev reflected into a hole at lead $\beta$ and finally escapes
in lead $\gamma$. The threshold voltage for this process is
$V_{\beta\alpha}-V_{\gamma\beta}=2\Delta/e$. When $\gamma=\alpha$, we
recognize the usual MAR features for leads $\alpha$ and $\beta$ with
$n=2$, see Fig.~\ref{fig2}b.  When $\gamma\neq\alpha$, it determines a
new set of SGS lines at $V_{21}+V_{31}=\pm 2\Delta/e$,
$2V_{21}-V_{31}=\pm 2\Delta/e$, and $V_{21}-2V_{31}=\pm 2\Delta/e$
(see Fig.~\ref{fig2}c).

Similarly, the processes where an incoming electron undergoes two ARs
before escaping out at the upper gap edge in a S-lead determine
the standard set of MAR features with $n=3$, see Fig.~\ref{fig2}d, as
well as the lines: $V_{21}+2V_{31}=\pm 2\Delta/e$, $V_{21}-3V_{31}=\pm
2\Delta/e$, and $2V_{21}-3V_{31}=\pm 2\Delta/e$, plus three other
lines obtained by interchanging the indices 2 and 3 (see
Fig.~\ref{fig2}e). The generalization to arbitrary numbers of ARs is
straightforward and yields the lines \refe{eq:MAR-SSS1}.

Now, we discuss the SGS lines given by \refE{eq:MAR-SSS2}.  These
lines, in contrast to those in \refE{eq:MAR-SSS1}, arise due to
the interplay of two different processes that take a quasiparticle
from below the gap in lead $\alpha$ to the upper gap edge in lead
$\beta$ or $\gamma$. In the absence of AR, for e.g.
$V_{21}=V_{31}>2\Delta/e$, an electron injected at energy $\eps<-\Delta-eV_{1}$ in
lead 1 can be directly transmitted through the structure, out into
lead 2 or 3 (see Fig.~\ref{fig2}f). However, for $V_{21}>V_{31}$
($V_{31}>V_{21}$) there is an energy interval
$\Delta-eV_{2}<\eps<\Delta-eV_{3}$ ($\Delta-eV_{3}<\eps<\Delta-eV_{2}$) for
which injected electrons only can escape out into lead 2 (3). This
opening up of one and closing of another transport process is the origin of the SGS-line
at $V_{21}=V_{31}$, with $|V_{21}|>2\Delta/e$. Similarly, we have
SGS-lines at $V_{31}=0$ with $|V_{21}|>2 \Delta/e$, and $V_{21}=0$ with $|V_{31}|>2
\Delta/e$, due to the interplay between processes with no AR.

For one AR in one of the processes we find in the same way SGS-lines
at $V_{21}=2V_{31}$ with $|V_{21}|>2\Delta/e$, $V_{31}=2V_{21}$ with $|V_{31}|>2\Delta/e$ 
and $V_{21}=-V_{31}$ with $|V_{21}|>\Delta/e$. If
in total two ARs in the two processes are considered we find first the
same SGS-lines as for zero AR but with a voltage threshold at
$\Delta$. Second, we also have the line at $2V_{31}=3V_{21}$
with $|V_{21}|>\Delta/e$ (see
Fig. \ref{fig2}g) plus five more lines obtained by permutation of the
indices 1, 2 and 3. Again, the generalization to arbitrary numbers of ARs
yields the lines in Eq. (\ref{eq:MAR-SSS2}).

On the whole, the SGS lines are determined by \refE{eq:MAR-SSS} and
they only get smeared at temperatures $kT\sim\Delta$, close to the
superconducting critical temperature of the leads. We also expect that
a good contact between the S-leads and the intermediate N-region is
required to observe many lines. For poor interfaces, AR is strongly
suppressed and only the lines corresponding to a small number of ARs
are visible. We stress that Eq.~(\ref{eq:MAR-SSS}) is quite
generic; it holds for any incoherent as well as short, coherent
junction. For long, coherent junctions, i.e. with normal-state dwell
time comparable to $\hbar/\Delta$, additional SGS features related to
the inverse dwell time (Thouless energy) are expected
\cite{samuelsson:2002,cuevas:2006}. The appearance of the
SGS-features -- steps, peaks/dips or cusps in the voltage dependence of 
the differential conductances or second-order derivatives -- depends on the transparencies
of the NS-interfaces and on the model for electron propagation in the
N region \cite{Naz}.

In the following, we investigate the SGS-lines and the MAR-transport
in a three-terminal normal diffusive metal contacted to the S leads
through barriers with arbitrary transparencies (see
Fig.~\ref{fig1}b). We restrict ourselves to the incoherent case: The
length of the normal arms is assumed to be much larger than a
pair-breaking length set either by magnetic impurities or by an
external magnetic field. Then, the coherence between an incident
electron and Andreev reflected hole does not extend into the normal
metal and no non-dissipative Josephson current can flow through the
structure. In particular, the electronic properties of the normal
metal are fully described by the energy-resolved nonequilibrium
distribution functions for electrons and holes.

The current flowing through each arm, 
\beq
\label{eq:current}
I_\alpha=\frac{e}{h}\int d\eps 
\left[i_\alpha^e(\eps)-i_\alpha^h(\eps)\right],
\eeq
is decomposed into spectral currents for electrons,
\beqa
\label{eq:spectral-current}
i_\alpha^e(\eps)
&=&
{\cal T}_\alpha(\eps+eV_\alpha)
[f^e(\eps)-f_0(\eps+eV_\alpha)]
\nonumber\\
&&
+
{\cal A}_\alpha(\eps+eV_\alpha)
[f^e(\eps)-f^h(\eps+2eV_\alpha)],
\eeqa
and holes, $i^h_\alpha(\eps)=-i^e_\alpha(-\eps)$.  Current
conservation at the connection point of the three arms, called the
central node, imposes
\beq
\label{eq:current-conservation}
\sum_\alpha i^{e}_\alpha(\eps)=0, \hspace{0.5cm}\sum_\alpha i^{h}_\alpha(\eps)=0.
\eeq
Here, $f^e(\eps)$ and $f^h(\eps)=1-f^e(-\eps)$ are the distribution
functions for electrons and holes in the central node, $f_0$ is the
Fermi distribution function at temperature $T$, and ${\cal T}_\alpha$
and ${\cal A}_\alpha$ are the coefficients for normal transmission and
AR between the lead $\alpha$ and the node. We further assume that the
barriers between the leads and the arms are point contacts
characterized by a number of channels $N_\alpha$ and a transparency
$\Gamma_\alpha$ in the normal state, with conductance
$G_{B\alpha}=(2e^2/h)N_\alpha\Gamma_\alpha$.  Then,
\begin{subequations}
\label{eq:coef-transmission}
\beqa
{\cal T}_\alpha(\eps)
&=&
\frac{N_\alpha T_\alpha(\eps)g_{D\alpha}}
{g_{D\alpha}+ N_{\alpha}T_\alpha(\eps)},
\\
{\cal A}_\alpha(\eps)
&=&
\frac{N_\alpha A_\alpha(\eps)g_{D_\alpha}^2}
{[g_{D\alpha}+ N_{\alpha}T_\alpha(\eps)]
[g_{D\alpha}+ N_{\alpha}\{T_\alpha(\eps)+2A_\alpha(\eps)\}]}. \nonumber \\
\eeqa
\end{subequations}
Here, $G_{D\alpha}=(2e^2/h)g_{D\alpha}$ are the conductances of the
normal arms and $T_\alpha(\eps)$ and $A_\alpha(\eps)$ are the normal-transmission
and AR probabilities at a single
NS-interface~\cite{blonder:1982}. \refE{eq:coef-transmission} allows
to describe the crossover between two limiting cases: When the
diffusive wire is in good contact with the leads, $G_{D\alpha}\ll
G_{B\alpha}$, they reduce to ${\cal
T}_\alpha(\eps)=g_{D\alpha}\theta(|\eps|-\Delta)$ and ${\cal
A}_\alpha(\eps)=[g_{D\alpha}/2]\theta(\Delta-|\eps|)$
~\cite{nagaev:2001}.  On the other hand, when $G_{D\alpha}\gg
G_{B\alpha}$, they describe an incoherent chaotic or diffusive dot
attached to the leads through point contacts~\cite{samuelsson:2002}.

In order to calculate the currents $I_{\alpha}$ in \refe{eq:current}
at given bias voltages, we need the distribution functions $f^e(\eps)$
and $f^h(\eps)$ in the central node. They are determined by an
infinite system of linear equations
\refe{eq:spectral-current}-\refe{eq:current-conservation} relating
them at various energies. In general, we solve this system numerically
with a standard Jacobi algorithm after projection on a discrete energy
grid. The convergence is very fast (only a few iterations are
necessary), yielding strongly nonequilibrium distribution functions
with sharp peaks and dips at energies in and around the gapped
region. The differential conductances $G_{\alpha\beta}=\partial
I_\alpha/\partial V_\beta$ are then obtained by numerical
differentiation.

We illustrate the results for a chaotic dot symmetrically contacted to
the leads. The SGS lines \refe{eq:MAR-SSS} in the differential
conductance are clearly visible in Fig.~\ref{fig1}a,c. For clarity, we
display only the result for positive values $V_{21}, V_{31}\geq
0$. The SGS lines appear in any local and nonlocal conductance and are
robust for temperatures up to the critical temperature of the leads
(not shown).
 
In addition to the SGS-lines we investigate the excess current and 
the conductance at low bias. Standard methods for MAR-transport at 
large~\cite{blonder:1982} or small voltage~\cite{nagaev:2001b} in 
two-terminal junctions are readily generalized to the multiterminal 
geometry considered here. We provide the results in the following.
At large bias voltages
$|V_{21}|,|V_{31}|,|V_{32}| \gg \Delta/e$ the currents flowing through
the leads are close to their normal state value:
\beq
\label{eq:G-normal}
I_\alpha^N
=
\sum_{\beta\neq\alpha}
\frac{G_\alpha G_\beta}{G_\Sigma}
(V_\alpha-V_\beta),
\eeq
where $G_\alpha^{-1}=G_{B\alpha}^{-1}+G_{D\alpha}^{-1}$ and
$G_\Sigma=\sum_\alpha G_\alpha$.  The effect of the S leads is to
induce an excess current that takes a finite value at large voltages,
$I_\alpha=I_\alpha^N+I_\alpha^\mathrm{exc}$, where for $kT\ll \Delta$
\beq
\label{eq:exc}
I_\alpha^\mathrm{exc}
=
\frac{\Delta}{2e}
\sum_{\beta,\gamma,\eta} 
\left(\delta_{\beta\alpha}-\delta_{\gamma\alpha}\right)
\frac{G_\beta\Lambda_\gamma G_\eta}{G_\Sigma}\mbox{sign}(V_{\eta\gamma}),
\eeq
and
\begin{equation}
\Lambda_\alpha=\int\frac{dE}{\Delta}\frac{{\cal T}_\alpha(E)+2{\cal
A}_\alpha(E)-hG_\alpha/(2e^2)} {{\cal T}_\alpha(E)+2{\cal
A}_\alpha(E)+h(G_\Sigma-G_\alpha)/(2e^2)}.
\end{equation}
When the resistance of the interfaces is negligible ($G_{D\alpha}\ll
G_{B\alpha}$), we obtain $\Lambda_\alpha=0$ and thus
$I_{\alpha}^\mathrm{exc}=0$. However, in the general case the excess current
remains finite. In the case of a symmetric device with $G_\alpha\equiv
G$ and $\Lambda_\alpha\equiv \Lambda$ all identical, the excess
current \refe{eq:exc} reduces to
$I^\mathrm{exc}_\alpha=I_0\sum_\beta\mathrm{sign}(V_\alpha-V_\beta)$
where $I_0=G\Lambda\Delta/(2e)$.  Thus it can only take one of the
values $I^\mathrm{exc}_\alpha=0,\pm 2I_0$, depending on the bias
voltages.

At energies inside the gap, $|\eps|<\Delta$, the coefficient for
normal transmission vanishes, ${\cal T}_\alpha=0$. Equations
\refe{eq:spectral-current}-\refe{eq:current-conservation} at vanishing
bias then reduce to a diffusion equation in energy space:
\begin{equation}
\frac{\partial}{\partial \eps}
\left({\cal D}(\eps)
\frac{\partial f^e(\eps)}{\partial \eps}
\right)=0
\label{endiff}
\end{equation}
where ${\cal D}(\eps) = \sum_{\alpha\beta}(V_\alpha-V_\beta)^2{\cal
A}_\alpha(\eps){\cal A}_\beta(\eps) /\sum_{\alpha}{\cal
A}_\alpha(\eps)$. Outside the gap, $|\eps|>\Delta$, $f^e(\eps)=f_0(\eps)$, 
giving boundary conditions $f^e(-\Delta)=1$ and $f^e(\Delta)=0$ at $kT \ll
\Delta$. Eq.~(\ref{endiff}) is then readily solved and
we find the current through lead $\alpha$:
\begin{equation}
I_\alpha=
\frac{e^2}{h}
\left(
\int_{-\Delta}^{\Delta}\frac{d \eps}{ {\cal D}(\eps)}
\frac{\partial {\cal D}(\eps)}{\partial V_\alpha}
\right)
/\int_{-\Delta}^{\Delta}\frac{d\eps}{ {\cal D}(\eps)}
\end{equation}
For a diffusive region contacted to the leads through perfectly
transparent interfaces (all $\Gamma_\alpha=1$), the coefficients
${\cal A_\alpha}$ are constant inside the gap~\cite{blonder:1982} 
and one finds:
\beq
\label{eq:G-lowV}
I_\alpha^{\mathrm{low}\,V} = \sum_{\beta\neq\alpha} \frac{\tilde
G_\alpha \tilde G_\beta}{\tilde G_\Sigma} (V_\alpha-V_\beta), 
\eeq
where $\tilde G_\alpha^{-1}=(2G_{B\alpha})^{-1}+G_{D\alpha}^{-1}$ and
$\tilde G_\Sigma=\sum_\alpha \tilde G_\alpha$.  \refE{eq:G-lowV}
reduces to the normal state result \refe{eq:G-normal} at
$G_{B\alpha}\gg G_{D\alpha}$, while it predicts local and nonlocal
conductance doubling compared to the normal state result at
$G_{B\alpha}\ll G_{D\alpha}$.  For a symmetric structure with
identical arms (arbitrary transparencies), we again obtain an
expression similar to \refe{eq:G-lowV} but with (all identical)
effective conductances: $\tilde G_\alpha = (4e^2/h)/
\int_{-\Delta}^{\Delta} d\eps/[2\Delta {\cal A}_\alpha(\eps)]$.

From these analyses we find that for a diffusive structure
with negligible interface resistances, the currents flowing through
it at $kT \ll \Delta$ coincide with their normal state
values both in the low and high voltage regions (no excess
current). Numerical calculations for intermediate voltages 
and temperatures up to $kT \sim \Delta$ also show
no SGS line. Therefore, the MARs neither show up in the
local~\cite{nagaev:2001b} nor in the nonlocal conductances.

Let us now compare the results with those for hybrid multiterminal
junctions with two S leads (bias voltages $V_2$ and $V_3$) and one N
lead (bias voltage $V_1$). There is again a complex pattern of
SGS lines at voltages:
\begin{subequations}
\label{eq:SSN}
\beqa
&&n(V_{21}-V_{31})=2  \Delta/e
\qquad (n\, \mathrm{integer}),
\label{eq:SSN1}
\\
&&nV_{21}-(n-1) V_{31}=\pm\Delta/e,
\label{eq:SSN2}
\\
&&V_{21}=V_{31} \quad \mathrm{at} \quad |V_{21}|>\Delta/e.
\label{eq:SSN3}
\eeqa
\end{subequations}
The lines \refe{eq:SSN1} are related to standard MARs between the S
leads, surviving up to the superconducting critical temperature. In
contrast, the lines \refe{eq:SSN2} arise when an electron is injected
from the N lead, at the chemical potential, and is emitted at the gap
edge of one of the S leads after several ARs between them. 
Since these lines are dependent on the position
of the chemical potential in the N lead, they are smeared already at
finite temperatures $kT \ll \Delta$. The last line \refe{eq:SSN3}
arises due to the interplay of two processes where a particle
emitted from below the chemical potential in the N lead reaches the
upper gap edge in one of the two S-leads. At zero temperature, this
process has a voltage threshold at $|V_{21}|>\Delta/e$. At finite
temperature $kT\ll\Delta$, the threshold is shifted toward lower
voltage $|V_{21}|\gtrsim(\Delta-kT)/e$ due to the smearing of the
distribution function near the chemical potential in the N lead.  It
eventually shades away at $kT\sim \Delta$.

\begin{figure}
\includegraphics[width=0.9\linewidth]{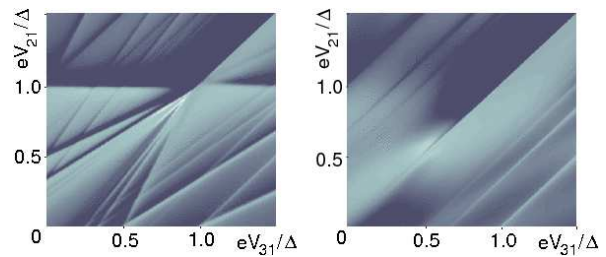}
\caption{Top view (similar to Fig. \ref{fig1}) of the differential
conductance $G_{21}$ for lead 1 normal and 2,3
superconducting. The conductances $G_{B\alpha}$ and $G_{D\alpha}$ 
are the same as in Fig. \ref{fig1} and temperatures are (a) $kT=0.01\Delta$ and (b)
$kT=0.15\Delta$.}
\label{fig3} 
\end{figure}

The set of equations
\refe{eq:spectral-current}-\refe{eq:current-conservation} can also be
used to determine the currents in an incoherent multiterminal junction
with N and S leads ($\Delta\rightarrow 0$ in N leads). The lines
\refe{eq:SSN} in the differential conductances and their temperature
dependence are clearly visible in Fig.~\ref{fig3}.

Thus we have formulated a theory for charge
transport in the incoherent three-terminal structure shown in
Fig.~\ref{fig1}. It could be easily generalized to a circuit theory 
for an arbitrary number
of leads, incoherent nodes and connectors between them. Other
connectors than the point contact and incoherent diffusive wire in
series considered here could also be addressed. For instance, the
quantum coherence of the AR process in vicinity of the S leads would
lead to renormalized values of the coefficients $T_\alpha(\eps)$ and
$A_\alpha(\eps)$, see Ref.~\cite{bezuglyi:2001}. Our approach could
also be extended to calculations of the current
correlations \cite{duhot:2009} and the full counting statistics
\cite{FCSincoh}.

In conclusion, we have addressed nonlocal currents in hybrid
multiterminal junctions. We have shown that multiple Andreev
reflections involving several leads give rise to a much richer
subharmonic gap structure than previously investigated in two-terminal
junctions. An experimental test of our predictions is
feasible in presently available superconducting systems.

We thank Yu. Nazarov for discussion.
MH acknowledges support from ANR-07-NANO011 ELEC-EPR and PS support
from the Swedish VR.


\begin{thebibliography}{99}
		
\bibitem{andreev:1964}
	A.F. Andreev,
	Sov. Phys. JETP \textbf{19}, 1228 (1964).

\bibitem{tinkham:1982} 
	T. M. Klapwijk, G. E. Blonder, and M. Tinkham, 
	Physica B \and C \textbf{109-110}, 1657 (1982).

\bibitem{octavio:1983} 
	M. Octavio {\it et al.},
	Phys. Rev. B \textbf{27}, 6739 (1983).

\bibitem{pierre:2001}
	F. Pierre {\it et al.},
	Phys. Rev. Lett. \textbf{86}, 1078 (2001).

\bibitem{Diel:1997}
	P. Dieleman et al., Phys. Rev. Lett. {\bf 79}, 3486 (1997).

\bibitem{metalnoise} 
	X. Jehl et al., Phys. Rev. Lett. {\bf 83}, 1660 (1999); 
	T. Hoss {\it et al.}, Phys. Rev. B {\bf 62}, 4079 (2000);
	C. Hoffmann {\it et al.}, Phys. Rev. B \textbf{70}, 180503 (2004).

\bibitem{Cron:2001}
	R. Cron et al., Phys. Rev. Lett. {\bf 86}, 4104 (2001).

\bibitem{buitelaar:2003}
	M.R. Buitelaar {\it et al.},
	Phys. Rev. Lett. \textbf{91}, 057005 (2003);
	P. Jarillo-Herrero, J. A. van Dam, and L. Kouwenhoven, Nature {\bf 439}, 953 (2006); 
	H.I. Jorgensen {\it et al.}, Phys. Rev. Lett. \textbf{96}, 207003 (2006).
	
\bibitem{xiang:2006}
	Y.J. Doh {\it et al}, Science {\bf 309}, 272 (2005);
	Jie Xiang {\it et al.}
	Nature Nanotechnology \textbf{1}, 208 (2006).

\bibitem{heersche:2007}
	H.B. Heersche {\it et al.},
	Nature \textbf{446}, 56 (2007).

\bibitem{kutchinsky:1999}
	J. Kutchinsky {\it et al.},
	Phys. Rev. B {\bf 56}, R2932 (1997).

\bibitem{lantz:2002}
	J. Lantz {\it et al.},
	Phys. Rev. B \textbf{65}, 134523 (2002).
	
\bibitem{duhot:2009}
	S. Duhot {\it et al.},
	Phys. Rev. Lett. \textbf{102}, 086804 (2009).

\bibitem{Pierre} 
	We note that in Ref. \cite{pierre:2001} a superconducting probe 
	was tunnel coupled to a diffusive SNS-junction, giving only a few 
	of the SGS-lines in our Eq. (\ref{eq:MAR-SSS}).
	
\bibitem{samuelsson:2002}
	P. Samuelsson {\it et al.},
	Phys. Rev. B \textbf{65}, 180514 (2002).
	
\bibitem{cuevas:2006}	
	J. C. Cuevas  {\it et al.}, Phys. Rev. B {\bf 73}, 184505 (2006).

\bibitem{Naz} For a recent discussion see e.g. 
	{\it Quantum transport}, 
	Yu. V. Nazarov and Ya. M. Blanter,
	Cambridge University Press (2009). 

\bibitem{blonder:1982} 
	G. E. Blonder {\it et al.},
	Phys. Rev. B \textbf{25}, 4515 (1982).

\bibitem{nagaev:2001}
	K. E. Nagaev and M. B\"uttiker, 
	Phys. Rev. B \textbf{63}, 081301 (2001).

\bibitem{nagaev:2001b}
	K. E. Nagaev, 
	Phys. Rev. Lett. \textbf{86}, 3112 (2001).	
	
\bibitem{bezuglyi:2001}
	E. V. Bezuglyi {\it et al.},
	Phys. Rev. B \textbf{62}, 14439 (2000).

\bibitem{FCSincoh}
	S. Pilgram and P. Samuelsson, Phys. Rev. Lett. {\bf 94}, 086806 (2005).

\end{thebibliography}
\end{document}